# Choke flutter instability sources tracking with linearized calculations

## Abstract
**Purpose** – The choke flutter is a fluid-structure interaction that can lead to the failure of fan or compressor blade in turbojet engines. In Ultra High Bypass Ratio (UHBR) fan, the choke flutter appears at part speed regimes and at low or negative incidence when a strong shock-wave chokes the blade to blade channel. Localization of main excitation sources and the understanding of the different work exchange mechanisms will help to avoid deficient and dangerous fan design.
**Design/methodology/approach** - In this paper, an UHBR fan is analyzed using a time-linearized Reynolds-Averaged Navier-Stokes equation solver to investigate the choke flutter. The steady state and the imposed vibration (inter blade phase angle, reduced frequency and mode shape) are selected to be in choke flutter situation. Superposition principle induced by the linearization allow to decompose the blade in numerous small subsections to track the contribution of each local vibration to the global damping. All simulations have been performed on a 2D blade to blade extraction.
**Findings** - Result analysis points to a restricted number of excitation sources at the trailing edge which induce a large part of the work exchange in a limited region of the airfoil. Main phenomena suspected are the shock-wave motion and the shock-wave/boundary layer interaction.
**Originality/value** – An original excitation source tracking methodology allowed by the linearized calculation is addressed and applied to a UHBR fan test-case.
**Paper type** Research paper

## Nomenclature
| | | |
|---|---|---|
| Nn | = | Nominal speed |
| UHBR | = | Ultra High Bypass Ratio |
| $\dot{m}$ | = | Mass flow |
| $S_i$ | = | Moving node area |
| $\mathcal{U}$ | = | Maximal vibrating kinetic energy |
| $\mathcal{W}$ | = | Work extracted |
| $\pi_t$ | = | Pressure ratio |
| $\zeta$ | = | Damping coefficient |

## 1. Introduction
The choke flutter can lead to the failure of fan or compressor blade in turbojet engines. Choke flutter appears when a strong shock-wave chokes the blade to blade channel. In Ultra High Bypass Ratio (UHBR) fan, choke flutter appears at part speed regimes (typically 80 % of the nominal speed, Nn) and at low or negative incidence (high mass flow, low total pressure ratio). The steady flow is subsonic upstream and downstream of the blade row and supersonic in the blade to blade channel. A strong shock-wave chokes the channel from the suction side to the pressure side.

At present time, the scientific community does not agree to one common explanation to understand the physical mechanisms leading to choke flutter (Clark et al., 2004) but the shock-wave motion itself, the flow separation induced and the acoustic blockage seems to be determinant. The blade vibrations lead to the oscillation of the shock-wave. This one induces a dynamic loading of the structure which can lead to an aeroelastic instability. The shock-wave motion is known to highly contribute to the aeroelastic behavior of the blade (Micklow and Jeffers, 1981). In flutter the only source of unsteadiness is the blade motion itself and differs from the transonic buffeting linked to a phase locking between two aerodynamics phenomena. If strong enough, the shock-wave can interact with the boundary layer and induce the separation of the flow. The relative importance of the flow separation and the shock-wave motion seems to be case dependant. Some publications conclude to a destabilizing effect for the oscillation of the shock-wave and a stabilizing effect of the flow separation (e.g. Isomura and Giles, 1998) and other conclude to the opposite results (e.g.Vahdati et al., 2001).

The acoustic blockage corresponds to backward travelling pressure waves generated downstream of the shock-wave and propagating upstream. When reaching the shock-wave, the velocity of backward travelling pressure waves decreases which leads to an increase of their amplitude (Atassi et al., 1995). For choke flutter, previous studies have shown the important contribution of the acoustic blockage (Ferrand, 1987; Rendu et al., 2016).

Computational Fluid Dynamics (CFD) is usually the only affordable way to obtain a time and space resolved flow field to investigate choke flutter physical mechanisms. In turbomachinery, the blade stability is generally obtained through the energetic method (Marshall and Imregun, 1996). This method relies on the radial decomposition of the 3D blade in a sum of 2D airfoils. The damping coefficient is computed on each 2D airfoil and the overall damping coefficient is obtained by an integral along the radius, from hub to tip. It is widely known, because the velocity and pressure fluctuations are the largest, that the region close to the tip gives the main contribution to the global damping coefficient.

To investigate the physical mechanisms leading to this instability, a specific test case of fan is selected and analyzed using a time-linearized Reynolds-Averaged Navier-Stokes equation solver. The mode shape, the interblade phase angle (IBPA) and the reduced frequency have been set to obtain a negative damping coefficient and so a case with a choke flutter instability. All simulations have been performed on a 2D blade to blade extraction at 90% height. Based on methods developed in previous study (Rendu et al., 2017), the contribution to the global damping coefficient induced by the vibration at each surface mesh node can be determined, thanks to the superposition principle. The studied case corresponds to a representative design of an Ultra High Bypass Ratio (UHBR) fan. This first fan design, named ECL5v1, is a part of a larger ongoing project on the numerical and experimental investigations of aeroelastic and aerodynamic instabilities at Ecole Centrale de Lyon. The presented vibration decomposition methodology has been already applyed to parametric investigation of the choke flutter on ECL5v1 case (Duquesne, Aubert, et al., 2018a, 2018b; Duquesne, Rendu, et al., 2018). In this paper, the ECL5v1 case is used to illustrate the typical results obtained with the decomposition methodology.

## 2. Numerical methods
### 2.1. Steady RANS solver
The compressible RANS solver Turb'Flow is used in this work to compute the 2D steady flow in a 90% height blade to blade channel. This solver relies on vertex centred finite volume method on multi-block structured grids (Smati et al., 1998). Convective fluxes are obtained through upwind scheme of Roe (Roe, 1981) with Monotonic Upstream-centred Scheme for Conservative Laws (MUSCL) interpolation of third order (van Leer, 1979). The interpolation order is reduced in strong gradient zones according to Harmonic Cubic Upwind Interpolation (H-CUI) limiter (Waterson and Deconinck, 2007). Diffusive fluxes are obtained through central interpolation of conservative variables. The pseudo time discretization relies on backward Euler with CFL=20 and local time step to speed up the convergence. The linear problem arising from the implicit method is solved through GMRES iterative method (Saad and Schultz, 1986). The flow is considered fully turbulent and the k-ω turbulence model of Wilcox (Wilcox, 1988) has been used. At the wall ω-value is extrapolated to be assumed infinite (Menter, 1993).

The Linearized RANS (LRANS) solver Turb'Lin is used to compute the harmonic flow around the steady state. This solver has been previously validated on transonic separated flows and the turbulence model has also been linearized because of the separated flows (Philit et al., 2012; Rendu et al., 2015). The solution is obtained in the frequency domain by solving the linear system. Spatial discretization relies on Jameson et al. (Jameson et al., 1981) centred scheme with linearized pressure sensor.

### 2.2. Aeroelasticity

The complex amplitude of displacement $\widetilde{\delta x}$ and velocity $\widetilde{V}$ are imposed at each node of the blade mesh to model the blades oscillation. The steady position of the blade is chosen as the phase origin. This yields

$$R(\widetilde{\delta x}) = 0 \quad ; \quad I(\widetilde{V}) = 0 \quad (1)$$

The interblade phase angle (IBPA) $\sigma$ is modelled through quasi-periodic boundary conditions in azimuthal direction

$$\tilde{q}(x_b + g) = \tilde{q}(x_b)e^{j\sigma} \quad (2)$$

where $\tilde{q}$ is the complex amplitude of conservative variable fluctuations, $x_b$ the domain boundary and g the interblade pitch, more details about periodic boundary layer in (Bakhle et al., 2004).

The work $\mathcal{W}$ extracted by the flow to the structure is written according to the convention of Verdon (Verdon, 1987). The damping coefficient is then obtained by the integral of the extracted work along the blade surface

$$\zeta = \frac{1}{4\pi} \frac{\iint_\Omega \mathcal{W} d\Omega}{\mathcal{U}} \quad (3)$$

where $\Omega$ is the fluid-structure contact interface and $\mathcal{U}$ the maximal vibrating kinetic energy. The work can be written as

$$\mathcal{W} = \int_0^T \left[-\widetilde{Ps}(x,t) * S(x,t)\right]^t \cdot \widetilde{V}(x,t) dt \quad (4)$$

where $\widetilde{Ps}$ is the instantaneous static pressure, S the vector associated to the instantaneous surface, oriented towards the structure, and $\widetilde{V}$ the instantaneous velocity vector associated to the blade displacement. In frequency domain, neglecting second order terms, the only contribution to the unsteady work is, for a rigid body motion,

$$R(\widetilde{Ps}^1)\tilde{S} \cdot R(\widetilde{V}^1) \quad (5)$$

where $\widetilde{Ps}^1$ and $\widetilde{V}^1$ are the complex amplitude of first harmonic of static pressure and velocity vector, respectively. Thus, only the real part of fluctuating static pressure contributes to the stability of the fluid-structure interaction.

### 2.3. Blade vibration decomposition

Superposition principle induced by the linearization of RANS equation leads to the equality between the unsteady flow generated by the vibration of the whole blade and the sum of the vibration of each surface mesh node. The blade vibration is modelled by imposing displacement and velocity on each node of the blade surface mesh. The blade vibration can thus be decomposed in an arbitrary number of zones $N$ and the global damping coefficient can be computed by the sum of the damping coefficient associated to each vibration. Formally,

$$\zeta = \sum_i^N \zeta_i \quad (6)$$

$$\zeta_i = \frac{1}{4\pi\mathcal{U}} \int \int_\Omega R(\widetilde{Ps_i^1}) \mathbf{S} \cdot R(\widetilde{V}^1) d\Omega \quad (7)$$

where $\widetilde{Ps_i}$ represents the pressure fluctuations generated by the motion of zone $i$.

To avoid even-odd decoupling, the blade is decomposed into pairs of adjacent mesh nodes. Each computation consists in the vibration of two adjacent nodes: first calculation is computed with the motion of the first and second mesh nodes, next calculation by the motion of the third and fourth nodes, etc.

The equality of the global damping coefficient, and its repartition along the chord, for the entire blade vibration and the result of the sum of the motion of each node has been checked. This decomposition strategy has some subtleties: the distance between adjacent nodes is not constant, so to compare the contribution of each nodes association the extracted work need to be normalized by the length between the two points. The blade meshing has an odd number of nodes; after some tests, the last segment is composed with the last three nodes instead of two. This choice has no consequence on the final result and the length of the last segment stays small due to the mesh high definition in this zone (the trailing edge). The set of calculations includes 424 L-URANS calculations (one by subsection). All these calculations are based on the same 2D steady state solution.

With such a vibration decomposition the total work exchange for the motion of only a single node can be determined around all the blade. Only the node can product pressure waves. The vibrating node can be interpreted as a *source*. By tracing the work along the chord for the vibration of one node (the source), zones where the work is significantly exchanged can be determined; these are referred by the term of *receptor* (of the pressure wave induced by the source). For example, a local source downstream a shockwave can produce regressive pressure waves which interact with the shockwave to exchange work at the wall receptor, normally in the vicinity of the shockwave.

## 3. Studied configuration
### 3.1. UHBR fan
The chosen test case is the Ultra High Bypass Ratio (UHBR) fan ECL5v1. The ECL5 design goals are to generate selected aeroelastic and aerodynamic instabilities, including the choke flutter at part-speed regime, and remains representative of future transonic UHBR fan. The operating range of the ECL5v1 fan, issues from numerical simulations, is plotted in figure 1 for three different rotational speeds (nominal speed Nn=10 450 rpm). The maximum isentropic efficiency, not shown here, varies between 90% and 95% depending on the rotational speed.

As already stated, the energetic method allows to decompose 3D blade in a sum of 2D airfoils. Most of the extracted work is generated close to the tip due to high levels of both blade velocity and pressure fluctuations. Therefore, a 2D blade to blade channel mesh has thus been extracted at 90% of ECL5v1 height to run the aeroelastic study. At this height the blade surface shows thin, highly staggered blades with low camber, which is typical of transonic fan tip airfoils.

Choke flutter is associated with negative incidence and strong shock-wave choking the interblade channel. It appears for part-speed regime, typically around 80% of the nominal rotational speed. For the aeroelastic study, the operating point showing the highest massflow on 80 Nn speed characteristic line is thus chosen (in blue in figure 1).

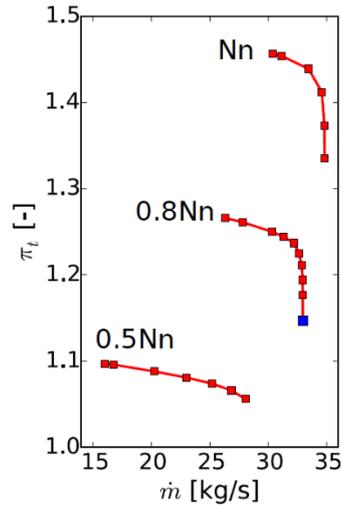

**Figure 1.** Pressure ratio ($\pi_t$) in function of mass flow ($\dot{m}$) for ECL5v1 - choked operating point in blue.

*3.2. Steady flow*
The mesh used for both steady and unsteady computations has been obtained through a convergence study. It consists in 106 007 points with $y^+ < 1$ for the first layer of cells close to the blade surface.

Total pressure, total temperature and azimuthal velocity are imposed at the upstream boundary and the static pressure at downstream boundary. The boundary conditions of the 2D-steady flow calculation are set to preserve the shock-wave position from the 3D calculation. The steady relative Mach number associated with the choked flow is plotted in figure 2.

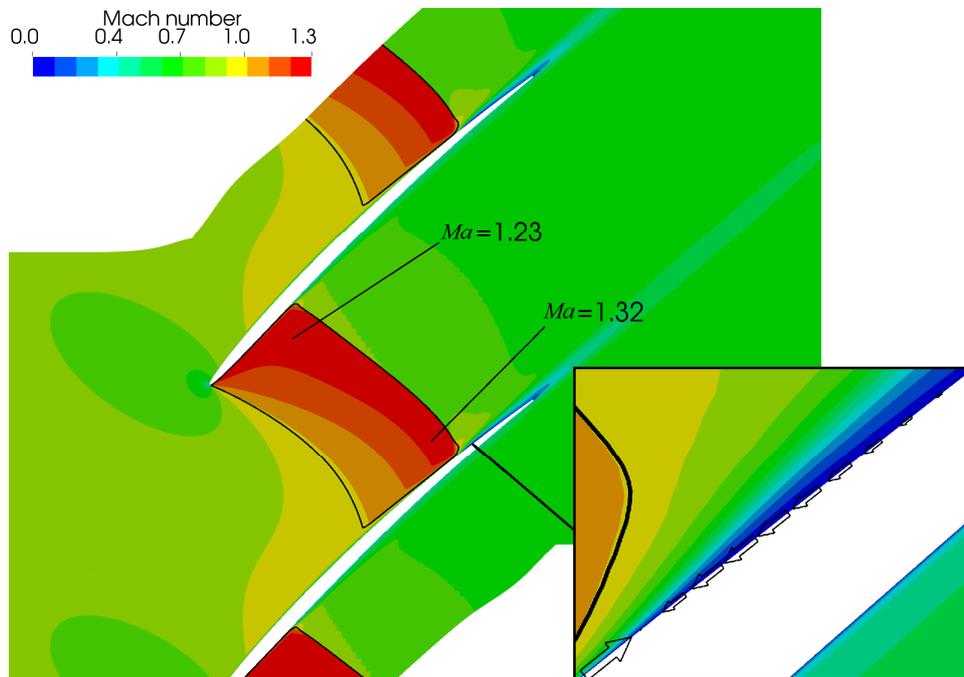

**Figure 2.** Steady relative Mach number for choked flow. Solid black lines represent the sonic lines (Mach number=1). On the enlargement, velocity vectors are plotted only on the suction side at the second mesh node.

Looking at the leading edge zone, negative incidence can be seen as well as a supersonic region choking the interblade channel and terminated by a strong shock-wave. On the pressure side, the maximal Mach number is 1.23 and the boundary layer is attached to the blade downstream of the shock-wave. On the suction side, the Mach number reaches 1.32 which leads to the separation of the boundary layer downstream of the shock-wave. The separation is closed and the reattachment point is located 8.3% of chord downstream of the separation point.

*3.3. Modeshape*
In this study, the chosen mode shape consists in a rotation of the airfoil around its leading edge without the deformation of the blade surface (i.e. a rigid body motion). This mode shape is representative of the first 3D torsion mode of the blade where the transonic flutter is observed. Motion of adjacent blades can present a phase shift called interblade phase angle or IBPA (frequency and mode shape remain identical between blades). The IBPA is by convention positive when the wave propagates in the same direction as the rotor speed and negative otherwise. The reduced frequency, for turbomachinery aeroelastic study, represents the ratio between the time of flight of a fluid particle along the chord and the time of a mechanical vibration period. In this work based on previous study, the IBPA is set at 90° and the reduced frequency is low at 0.15 (Rendu et al., 2017). The damping coefficient is negative. The work exchange is from the fluid to the blade, so this case presents a choke flutter instability.

A sketch of three adjacent blades position during the vibration cycle is plotted in figure 3. For each blade, colours correspond to different instants (-T/4,T,T/4). Vibration amplitude and interblade distant are modified for illustration purpose. The effective solidity (spacing/chord ratio) is 1.37. The out of phase blades vibration induces different passage section for adjacent interblade channels (see the same instant for the two channels in figure 3). This area fluctuation leads to strong velocity fluctuations.

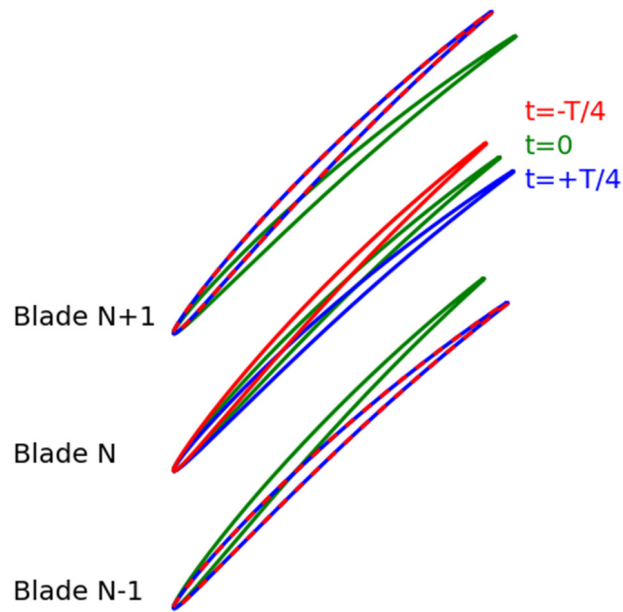

**Figure 3.** Sketch of the vibration of three adjacent blades at three different instants, airfoil colours show the different instants: -T/4,T,T/4. Vibration amplitude and interblade distant are modified for illustration purpose. For adjacent blades, blade positions at instants ±T/4 are superposed.

**4. Identifying main flutter source**

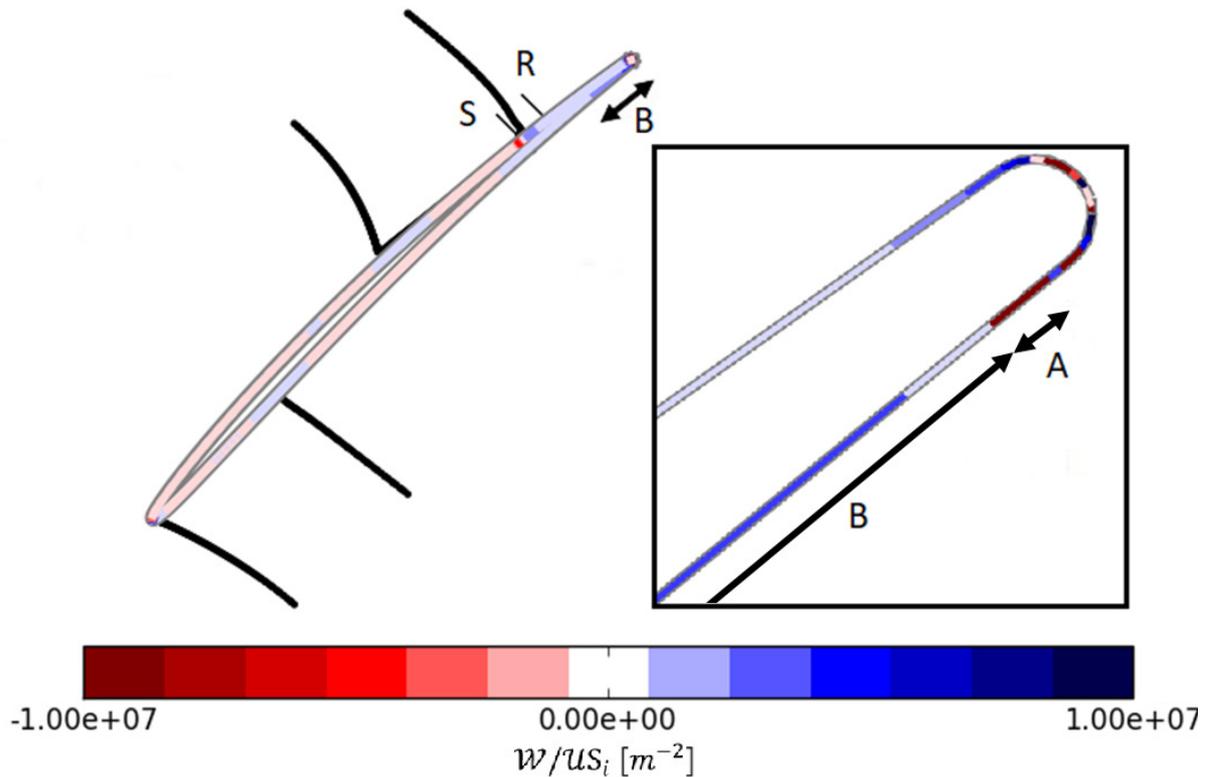

**Figure 4.** Normalized work on the airfoil, sonic line is in black, (S) and (R) are respectively separation and reattachment points, selected zones are referred by A,B. Grey zone corresponds to values near zero.

To determine the main flutter sources, the sum of the work along the chord is performed for the individual motion of each mesh point. One colored dot in figure 4 presents this sum induced by the motion of the mesh point located at the same position. Here, the work is normalized by the length of the moving point to determine main flutter sources without the effect of the associated mesh segment length. Colour-scale has been restricted for presentation purpose due to local high values near the trailing edge. In figure 4, the supersonic zone is delimited by black lines and the position of the separation point and reattachment point are reported by (S) and (R) respectively. The sub-figure is a trailing edge enlargement. Two zones have been selected because of their high contribution on stability in a restricted area. These zones are reported in figure 4 with a reference letter (A,B).

The A-zone is located on the pressure side near (but not on) the trailing edge. The vibration of this small zone, less than 0.1% of the total airfoil length, induces 56% of the negative (destabilizing) work. The vibration of the zone just upstream the A-zone on the pressure side (B-zone) induces a large part, 44%, of the positive (stabilizing) work. Even if B-zone is larger than the A-zone, it remains small (4.2% of the total airfoil length). The stabilizing work induced by the motion of the B-zone cannot compensate for the large amount of destabilizing work induced by the motion of the A-zone: the cumulative work of A and B-zones is negative (destabilizing). From these results, the last 5% at the end of the pressure side near the trailing edge seems leading the stability of motion of the overall blade.

## 5. Stabilizing/destabilizing areas induce by the main flutter sources

In the previous section, main excitation sources are localized. In this section downstream pressure side zone vibration (zone A and B) are selected to present the type of analysis provided by this method.

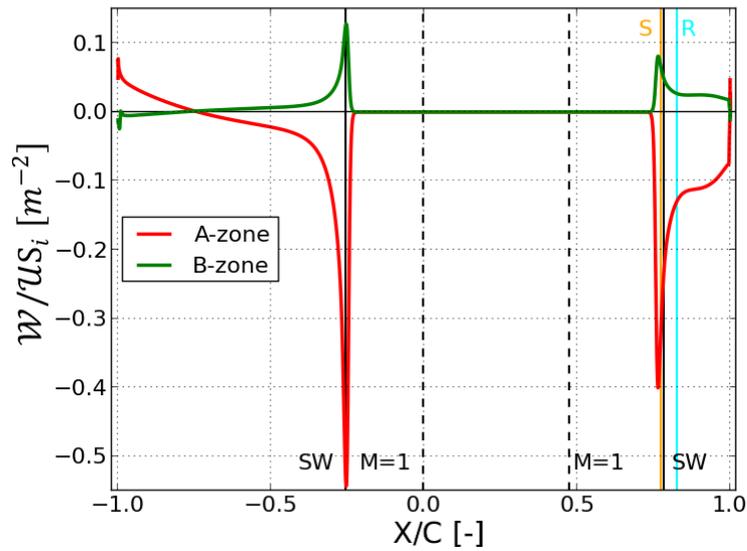

**Figure 5**. Work exchange along all the blade chord induced by the pressure wave from the motion of a point in A and B-zone (leading edge at X/C=0, pressure side: X/C<0, suction side: X/C>0). Moving point in A and B-zone are in red and green respectively. Separation, reattachment, sonic line and shock-wave positions are represented by vertical lines noted S, R, M=1 and SW respectively.

Figure 5 presents the work extracted locally by the blade for the motion of the most destabilizing point from A-zone (in red) and for the motion of the most stabilizing point from B-zone (in green). In this figure the work exchange is the result of the pressure fluctuation induced by the motion of one point in zone A or B and the motion of all the blade. For example, only the zone A vibrates so only this one can produce pressure waves (concept of source) but these waves can exchange work with all the blade (concept of receptor). The two selected points are close together and on the pressure side downstream the shock-wave near the trailing edge. According to previous results, the cumulative destabilizing work induced by the vibration in the A-zone is larger than the stabilizing work induced in the B-zone. In both cases the work exchange is located at or downstream the steady-state shock-waves position. The backward travelling pressure waves are stopped by the shock-wave and their important amplification at the shock-wave is compatible with an acoustic blockage. On the suction side the flow separation bubble seems to interact with the shock-wave motion to increase (versus the pressure side) the work exchange.

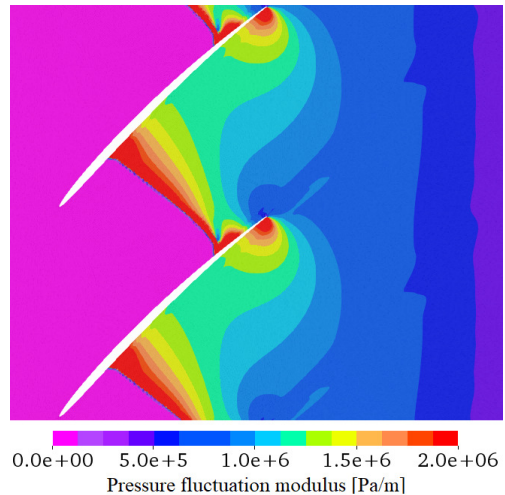

**Figure 6.** Pressure fluctuation modulus induced by the motion of the most destabilizing point in A-zone.

The work exchange repartition downstream the shock-wave can also be supported by the pressure fluctuation modulus mapping at figure 6. This figure presents the result (in red figure 5) for the motion of a point of the A-zone. The same trend can be observed for the motion of the point in B-zone but with a smaller amplitude (not shown here for brevity reason). The pressure fluctuations are generated by the motion of the point near the trailing edge and the pressure waves travel in both directions. Downstream, pressure waves are dampen in less than two chords. Upstream, they travel in the blade to blade channel up to the shock-wave where theirs amplitude increases. The pressure fluctuation gradient on the suction side is more intense and pressure fluctuations higher than in the pressure side. This is contra-intuitive with the located source on the pressure side; the explanation seems to be in the interaction with the flow separation bubble.

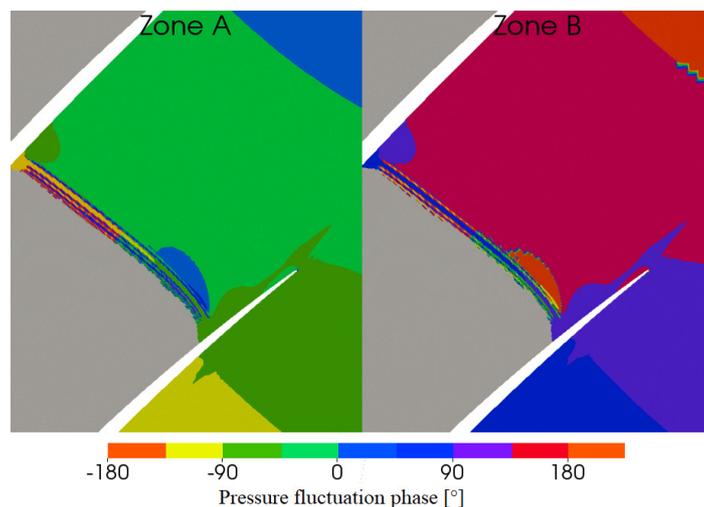

**Figure 7.** Pressure fluctuation phase induced by the motion of a point in A-zone at left and B-zone at right. Grey zone corresponds to zero fluctuation magnitude.

The figure 7 shows the pressure fluctuations phase mapping corresponding to the motion of the point in the zone near the trailing edge (A and B-zones). Grey zone corresponds to zero fluctuation magnitude, hence the phase is meaningless in this region. The phase mapping from the motion of a point in A-zone (destabilizing) and in B-zone (stabilizing, just upstream A-zone) have similar pattern but with a 180° phase shift. The phase shift is induced by the excitation source itself. This phase shift explains the opposite sign of the work exchange between motion of a point of the A-zone and a point of the B-zone.

## 6. Conclusions

The choke flutter in an Ultra High Bypass Ratio fan is analyzed using a time-linearized Reynolds-Averaged Navier-Stokes equation solver on a 2D blade to blade extraction at 90% height. The identification of the main sources of the work exchange between the flow and the blade and theirs effects on the stability of the entire blade have been performed using an innovating method based on the superposition principle. The contribution to the global damping coefficient induced by a local source is obtained with a simulation through only the vibration of a surface single mesh node.

Results have permitted to identify few zones with high work exchange values. The destabilizing cumulative work is associated to one main source. More than half of the destabilizing work is induced by the motion of a point source of excitation on the pressure side, downstream of the shock-wave near the trailing edge. The vibration of this zone induces backward travelling pressure waves, which propagates upstream up to the shock-wave.

The presented method can be used to locate and to investigate other excitation sources. CFD allows to perform huge parametric studies (turbulence modelling, nodal diameter, flutter frequency or mode shape) to develop a deep understanding of the flutter instabilities.